\documentclass{jltp}

\usepackage{graphicx}
\usepackage{dcolumn}
\usepackage{bm}
\usepackage[german,american]{babel}
\newcommand{\dwp}{DWP}

\newcommand{\figref}[1]{Fig.\ \ref{#1}}
\newcommand{\eqnref}[1]{Eqn.\ \ref{#1}}

\newcommand{\tls}{TLS}
\newcommand{\stm}{STM}

\newcommand{\peff}{\ensuremath{P_{\mathrm{eff}}}}
\newcommand{\ptot}{\ensuremath{P_{\mathrm{tot}}}}

\newcommand{\dmwrp}{\ensuremath{d_\mathrm{mwrp}}}

\newcommand{\bmlj}{BMLJ}
\newcommand{\un}[1]{\mathrm{#1}}

\title{How Cooperative are the Dynamics in Tunneling Systems?\\
       A Computer Study for an Atomic Model Glass.}

\author{J. Reinisch and A. Heuer}

\address{Westf\"{a}lische Wilhelms-Universit\"{a}t M\"{u}nster, Institut f\"{u}r Physikalische Chemie\\
and International Graduate School of Chemistry\\
Corrensstr. 30, 48149 M\"{u}nster, Germany}

\runninghead{J. Reinisch and A. Heuer}{How Cooperative are the Dynamics in Tunneling Systems?}

\begin{document}

\maketitle

\begin{abstract}
Via computer simulations of the standard binary Lennard-Jones
glass former we have obtained in a systematic way a large set of
close-by pairs of minima on the potential energy landscape, i.e.
double-well potentials (DWP). We analyze this set of DWP in two
directions. At low temperatures the symmetric DWP give rise to
tunneling systems. We compare the resulting low-temperature
anomalies with those, predicted by the standard tunneling model.
Deviations can be traced back to the energy dependence of the
relevant quantities like the number of tunneling systems.
Furthermore we analyze the local structure around a DWP as well as
the translational pattern during the transition between both
minima. Local density anomalies are crucial for the formation of a
tunneling system. Two very different kinds of tunneling systems
are observed, depending on the type of atom (small or large) which
forms the center of the tunneling system. In the first case the
tunneling system can be interpreted as a single-particle motion,
in the second case it is more collective.

\end{abstract}

\section{\label{introduction}INTRODUCTION}
Almost all kinds of disordered solids show anomalous physical
behavior at temperatures around 2 K \cite{Hunklinger:1976,Hunklinger:1986}.
Many of the observed
features can be explained by the Standard Tunneling Model (\stm)
\cite{Phillips:1972,Anderson:1972} and its generalization, the
Soft-Potential Model \cite{Karpov:1983,Buchenau:1991,Gil:1993,Parshin:1994}.
The basic idea of the \stm\ is to postulate the
existence of a broad distribution of Two Level States (\tls). A
TLS can be represented by a single degree of freedom, moving in a
double well shaped potential, thereby acting as a bistable mode.
The number of atoms, participating in this motion, does not enter
in the model. According to the \stm\ the distribution of \tls\ is
chosen as
\begin{equation}
 P(\Delta,\Delta_0)=P_0/\Delta_0,
\end{equation}
with an energy splitting
\begin{equation}
 E=\sqrt{\Delta^2+\Delta_0^2},
\end{equation}
where $\Delta$ is the asymmetry and $\Delta_0$ the tunneling
matrix element. The independence of $P_0$ from $E$ is an important
ingredient for the analytical predictions of the \stm. The \tls\
couple to strain and electric fields and therefore influence the
heat capacity, thermal conductivity, sound absorption, dielectric
response and other quantities; see the review by Phillips \cite{Phillips:1987}.

The \stm\ predicts a linear dependence of the heat capacity on
temperature and a quadratic dependence of the thermal conductivity
on temperature. The observed small deviations from the predicted
thermal conductivity behavior can be explained by an energy
dependence of $P_0$ and the deformation potential, describing the
coupling to strain \cite{HeuerSilbey:1993b}. Apart from these
minor deviations the \stm\ gives a good general agreement with
experimental results down to temperatures around 100 mK.

Below this temperature, however, deviations from the \stm\ occur
\cite{Natelson:1998,Classen:2000,Rosenberg:2003}. It is believed
that this behavior is caused by interacting \tls\ which are not
covered by the \stm. The strong change of the dielectric response
on an applied weak magnetic field which has been observed
\cite{Strehlow:1998,Ludwig:2002,Kettemann:1999} are also not
consistent with the \stm. As shown recently, these effects can be
understood by taking into account the interaction of the
quadrupole moments of the nuclei, involved in the dynamics, with
the local electric field gradient tensors \cite{Wuerger:2002}.

Whereas it is very difficult to obtain microscopic information
about the nature of TLS from experiments, such information can be
supplied by computer simulations
\cite{Trachenko:1998,Trachenko:2000,Vegge:2001,Schober:2002}.
Most bistable modes
will display an asymmetry much larger than 1 K. In general, energy
differences between adjacent energy minima in glass-forming
systems may range up to values much larger than $T_g$
\cite{Doliwa:2003a}. Therefore, modes with asymmetries less than
$T_g$ may be regarded as relatively symmetric, even if they do not
contribute to the low-temperature anomalies. We will denote them
as double-well potentials \dwp. In this manuscript we perform a
systematic search for DWP and analyze their microscopic
properties. This work extends previous work by Heuer and Silbey
\cite{HeuerSilbey:1993a,HeuerSilbey:1993b} in which circa 300
\dwp\ could be identified for a binary Lennard-Jones system. This
relatively small number of \dwp\ was already sufficient to obtain
direct information about the participation ratios and the absolute
number of DWP. In particular it was possible to extract the
properties of the DWP with asymmetries less than 1 K, i.e. the TLS.
In this way the deviations from the \stm, the dependence of P on
energy and its consequences could be elucidated. Due to advances
in computer technology and application of a new algorithm to
locate \dwp\ we are now able to obtain a much larger set of \dwp.
In a recent paper we have shown that this set of \dwp\ is not
hampered by computational limitations like very fast cooling
schedules as compared to the experiment
\cite{Reinisch_condmat:2004}. The goal of the present work is
twofold. First, we repeat our previous analysis to check the
predictions of the \stm. Second, we study the microscopic nature
of the \dwp\ in detail. This analysis requires a large set of
\dwp\ and was not possible before.

\section{\label{computational_details}COMPUTATIONAL DETAILS}
As a model system we have used a binary mixture Lennard-Jones
system with 80\% A-particles and 20\% B-particles (\bmlj)
\cite{Weber:1985,Kob:1995,Kob:1999,Broderix:2000,Doliwa:2003a}.
BMLJ is one of the standard glass-forming systems
with very good glass-forming properties.  The used potential is of
the type
\begin{equation}               V_{\alpha\beta}=4\cdot\epsilon_{\alpha\beta}[(\sigma_{\alpha\beta}/r)^{12}
    -(\sigma_{\alpha\beta}/r)^{6}]+(a +b\cdot r),
\end{equation}
with $\sigma_{\un{AB}}=0.8\sigma_{\un{AA}}$, $\sigma_{\un{BB}}=0.88\sigma_{\un{AA}}$,
$\epsilon_{\un{AB}}=1.5\epsilon_{\un{AA}}$,
$\epsilon_{\un{BB}}=0.5\epsilon_{\un{AA}}$, $m_{\un{B}} = 0.5 m_{\un{A}}$.
The simulation
cell is a cube with a fixed edge length according to the number of
particles and an exact particle density of $\un{D}=1.2$. Periodic
boundary conditions are used to minimize finite size effects and a
linear function $a +b \cdot r$ has been added to the potential to
ensure continuous energies and forces at the cutoff $r_{\un{c}}=1.8$. The
units of length, mass and energy are $\sigma_{\un{AA}}$, $m_{\un{A}}$,
$\epsilon_{\un{AA}}$, the time step within these units is set to
$0.01$.

Previously, this potential has been used to mimic NiP \cite{Weber:1985}
with
$^{62}$Ni and $^{31}$P, using $\sigma_{\un{AA}} =$
2.2\ {\AA}, the average mass per particle as 55.8 g/mol and
$\epsilon_{\un{AA}}=$ 7765 J/mol. With this choice T=1 in LJ-units
corresponds to 934 K. We just mention in passing that our density
is the standard density for the BMLJ system, but is circa 20\%
higher than the density used for the mapping on the NiP system
\cite{Weber:1985}.

We apply molecular dynamics simulations using the velocity
Verlet algorithm to equilibrate and generate a set of independent
configurations. After minimizing these configurations we are
systematically looking for nearby local energy minima and finally
check whether these minima are connected by a simple barrier of
first order with the respective starting minimum. Details of this
search algorithm can be found in \cite{Reinisch_condmat:2004}. We
have studied different system sizes ($N=65,\ 130,\ 195,\ 260$) to
identify possible finite size effects. For the \dwp\ properties,
reported in \cite{Reinisch_condmat:2004}, as well as for the
results in the present work no finite size effects are present.
The observation, that the structure of the \dwp\ is not size
dependent, even for our very small systems, is in accordance with
data for the incoherent scattering function and the radial
distribution function which already for $N=60$ are close to the
macroscopic limit \cite{Buechner:1999}. Furthermore it is known
from previous work \cite{HeuerSilbey:1993a,Reinisch_condmat:2004},
that the number of participating particles
is much smaller than the actual system size for the large majority
of observed \dwp. In what follows we restrict ourselves to system
sizes $N=65$ and $N=130$. The smaller system sizes have the
advantage that the efficiency of the \dwp\ location algorithm is
best \cite{Reinisch_condmat:2004} and to a good approximation a
complete set of \dwp\ is found, which enables us to obtain an
estimate for the density of \dwp.

Of particular importance is the determination of the saddles
between adjacent minima. The energy of the saddle is
one of the
major ingredients to calculate the tunneling matrix element. We
have employed an algorithm, recently developed for the analysis of
supercooled liquids above the glass transition \cite{Doliwa:2003a}.

Throughout this section we will use the following definitions for distances.
\begin{equation}
  d_{mw}^2(\vec{r}_1, \vec{r}_2)=\sum_i^N (d_{i, x}^2+d_{i, y}^2+d_{i, z}^2)\cdot\frac{m_i}{\bar{m}}
\end{equation}
\begin{equation}
  d_{mwrp}=d_{mw}(\vec{r}_1,\vec{r}_{trans. state})+d_{mw}(\vec{r}_{trans. state},\vec{r}_1)
\end{equation}
$d_{mw}^2(\vec{r}_1, \vec{r}_2)$ is the mass weighted distance
between two configurations, and $d_{mwrp}$ is the mass weighted
reactions path approximation between two minima. The mass
weighting has been introduced because the tunneling matrix element
$\Delta_0$ depends on the mass weighted distance. The generated
sets of \dwp\ have been truncated in their parameters to guarantee
the completeness of the search \cite{Reinisch_condmat:2004}.
The maximum distance \dmwrp\ is
0.8 and the asymmetry is limited to 0.5 if not stated otherwise.
Within this parameter range we have found 6522 \dwp\ from 10009
starting minima for the 65 particle system and 2911 \dwp\ from
3100 starting minima for $N=130$, which is lowered by an
inefficiency of the location algorithm for larger systems
\cite{Reinisch_condmat:2004}.
If not stated otherwise all
presented data are generated from systems fully equilibrated at
$T_{equil}=0.5$, which is slightly above the critical
mode-coupling temperature \cite{Doliwa:2003a} $T_c=0.45$.

\section{\label{results}RESULTS}
\subsection{\label{r2} Comparison with Experimental Findings}
\begin{figure}
  \includegraphics[width=\textwidth,clip]{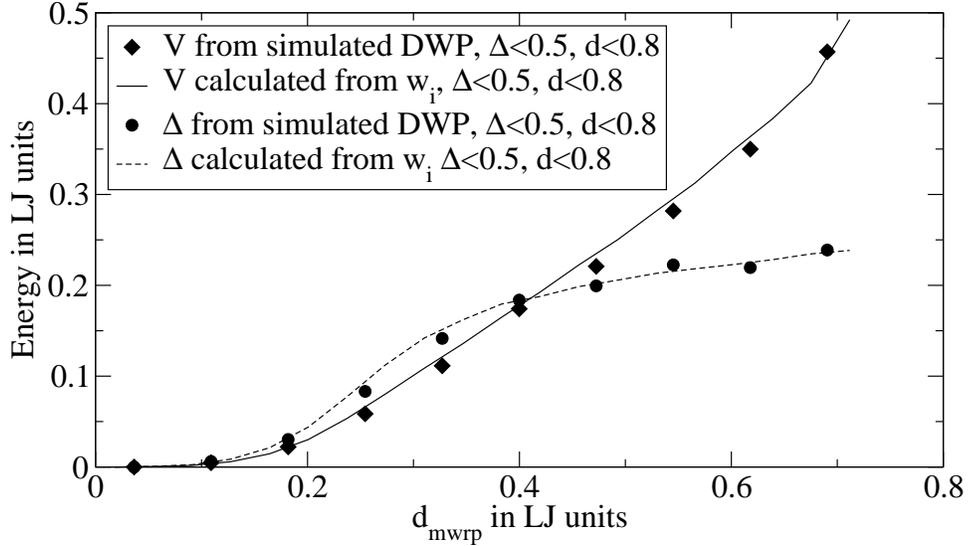}
  \caption{\label{asym_v_d_corr}Correlation of the averaged barrier height V
           and asymmetry $\Delta$ to the distance of the minima.
       Extra data are presented for comparison of the original data
       to those generated according to the $w_i$ distribution.}
\end{figure}
Our first goal is to calculate the density of \tls\ as expressed
by
\begin{equation}
  \label{def_peff}
  \peff(E)=\int_0^E d\Delta \int_0^E d \Delta_0
  \frac{\Delta_0^2}{E^2}\delta(E-\sqrt{\Delta_0^2+\Delta^2})P(\Delta,\Delta_0).
\end{equation}
 Within the \stm\ one has
\begin{equation}
  \peff(E)=P_0.
\end{equation}
The value of $\peff$ can be extracted from experimental data
\cite{Berret:1988}.

Every \dwp\ is characterized by the triplet $(\Delta,\dmwrp,V)$
where $V$ is the barrier height. There exist major statistical
correlations between all three quantities; see
\figref{asym_v_d_corr}. For example \dwp\ with small distances
between the minima naturally display small asymmetries and barrier
heights. Therefore direct extrapolation to nearly symmetric \dwp,
i.e. $\Delta < 1$ K, is not possible.  To solve this problem we
map every triplet $(\Delta,\dmwrp,V)$ on a triplet $(w_2,w_3,w_4)$
such that the \dwp, described by the fourth-order polynomial
\begin{equation}
  E_{pot}(x)=w_2*x^2-w_3*x^3+w_4*x^4 ,
\end{equation}
is characterized by the same triplet  $(\Delta,\dmwrp,V)$. This
procedure has been introduced in \cite{HeuerSilbey:1996,HeuerSilbey:1993a}.
Thus our set of \dwp\ is formally mapped on a
distribution $ P(w_2, w_3, w_4)$. For this mapping  each minimum
of the \dwp\ is taken as $x=0$, so that two sets of $w_i$ result
from one set of minima.

\begin{figure}
  \includegraphics[width=\textwidth,clip]{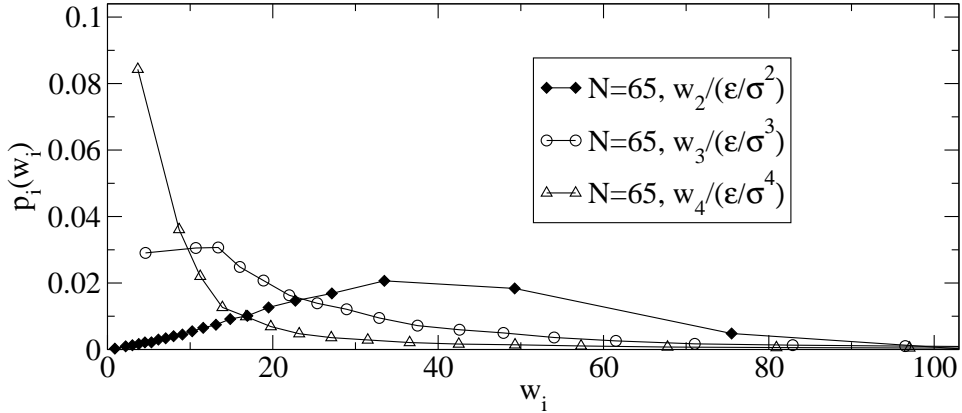}
  \caption{\label{et05w_distr}The $p_i(w_i)$ distributions as obtained from
  the \dwp\ for $N=65$. }
\end{figure}

Formally, the $w_i$ can be viewed as the Taylor-expansion
coefficients around the minima. In a disordered system the
individual terms are a sum of very different contributions from
the different pair interactions of the BJLM potential. Thus one
may speculate that the $w_i$ are statistically uncorrelated, i.e.
\begin{equation}
  P(w_2, w_3, w_4)=p_2(w_2)*p_3(w_3)*p_4(w_4).
\end{equation}
Via a least square fit in the range of parameters, in which a
systematic location of \dwp\ has been performed, the $p_i(w_i)$
have been determined. They are shown in \figref{et05w_distr}.
Via a statistical procedure, as outlined in \cite{HeuerArtikel},
we have checked that the $w_i$ are indeed uncorrelated.

Please note that the distribution $p_i(w_i)$ is very different to
the density $q_i(w_i)$ which denotes the number of \dwp\ with a
given value of $w_i$.  The reason is that the sub-volume of the
$(w_2,w_3,w_4)$-space which contains the \dwp\ in the given
parameter range, is highly non-trivial. Only in case that this
sub-volume were  a simple cube both distributions should be
identical.

Based on the distributions $p_i(w_i)$ it is possible to generate
\dwp\ with the correct statistics. Thus the generated \dwp\ should
have the same properties as the original \dwp. This is exemplified
in \figref{asym_v_d_corr} where we show the correlations of the \dwp\
three parameters for the original set of \dwp\ as well as for
the generated set of \dwp. No relevant deviations are present.
This underlines the validity of our assumption of uncorrelated
$w_i$.

\begin{figure}
  \includegraphics[width=\textwidth,clip]{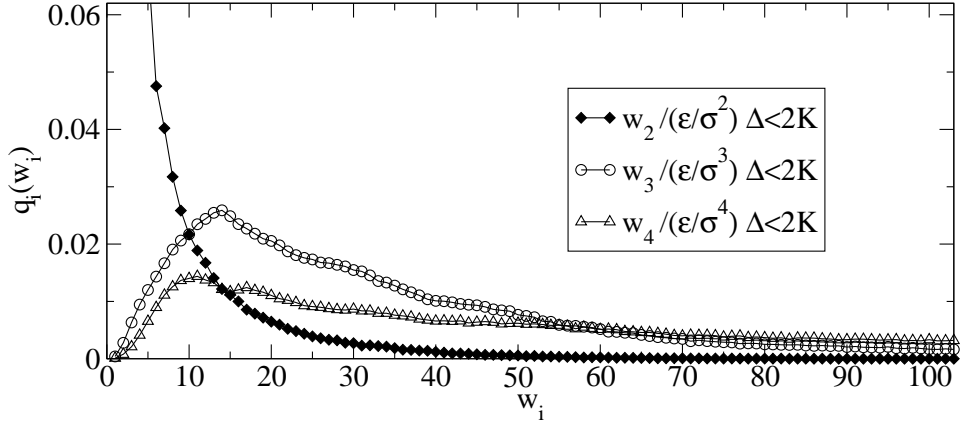}
  \caption{\label{et05art_w_distr_2K}The distributions $q_i(w_i)$ for $\Delta<2 \mathrm{ K}$
         as calculated from the observed \dwp.}
\end{figure}

Our main goal is to generate \dwp\ with very small asymmetry in
order to obtain information about typical \tls. For perfectly symmetric
\dwp\ one has
\begin{equation}
  w_2\cdot\ w_4/w_3^2=1/4.
\end{equation}
However this relation alone does not tell which range in the
$w_2,w_3,w_4$-space is relevant for nearly symmetric \dwp.
This question
is crucial, since from \figref{et05w_distr} it is obvious that the
$p_3$ and $p_4$ are ill-defined for small $w_3$ and $w_4$,
respectively. Therefore in a first step we have generated \dwp\
with $\Delta < 2$K. For this subset we have calculated the
$q_i$-distributions. They are shown in
\figref{et05art_w_distr_2K}. On their basis it is possible to
estimate which parameter range is essential for the generation of
nearly symmetric \dwp. Fortunately, it turns out that the range of
small $w_3$ and $w_4$ is irrelevant. Thus it is indeed possible to
use our original set of \dwp\ to generate symmetric \dwp, i.e.
\tls\ with the correct statistics. This shows that the range of
asymmetric \dwp\ contains sufficient information about nearly
symmetric \dwp. This justified our method to use change the
parameterization of the \dwp\ by using the $w_i$ rather than the
original parameters $(\Delta,\dmwrp,V)$.
\begin{figure}
  \includegraphics[width=\textwidth,clip]{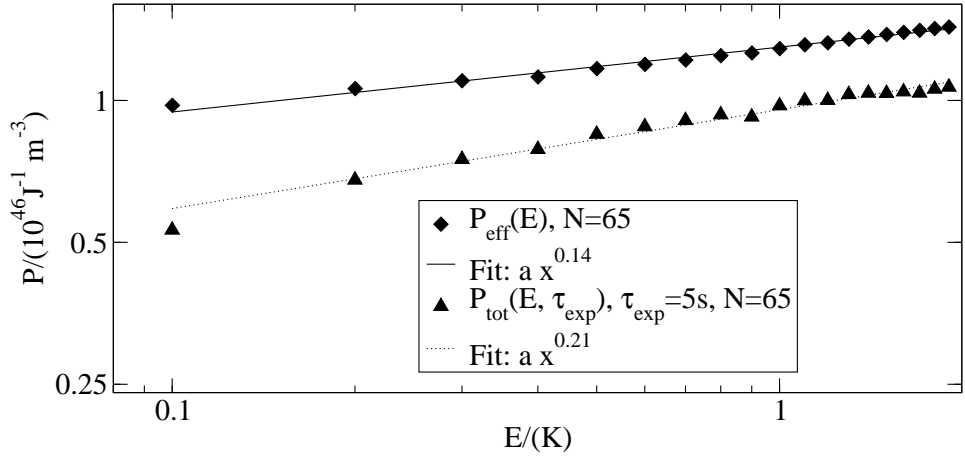}
  \caption{\label{pefftot} Energy dependency of \peff\ and $P_{tot}$.
           The data show the weighted and unweighted cumulative number of \tls\ per
       energy and volume.
              Both curves can roughly be approximated by a power law.}
\end{figure}

\begin{figure}
  \includegraphics[width=\textwidth,clip]{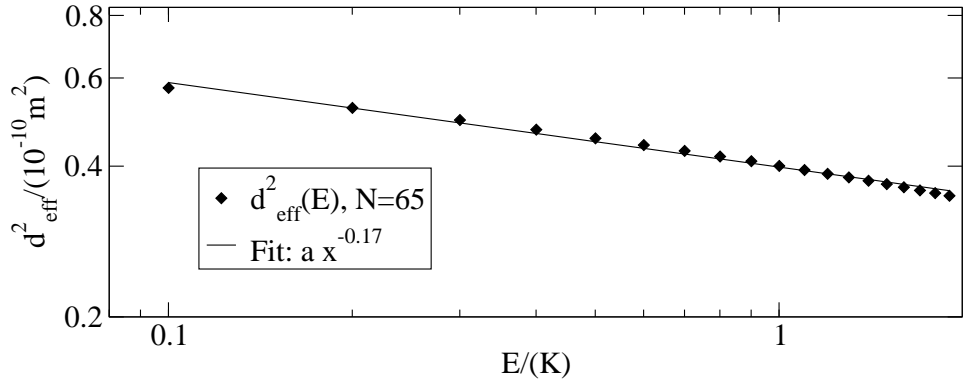}
  \caption{\label{d2eff} The effective distance decreases with higher energy splitting, because
                         large distances correlate with a high asymmetry and low $\Delta_0$.
             The curve can roughly be approximated by a power law.}
\end{figure}
For the \tls, generated in this way, we estimate $\Delta_0$ with
help of the Wentzel Kramer Brillouin (WKB) approximation. The
validity of the one dimensional WKB approximation is discussed in
\cite{Daldoss:1999} for the case of TLS in simulated argon
clusters. They observe only minor changes in the eigenfrequencies
during the transition, which makes the WKB approximation usable.
We make use of the WKB approximation in a slightly different form
than used by Phillips in \cite{Phillips:1987}:
\begin{equation}
  \Delta_0=E_0\cdot
  e^{-\frac{\pi}{4}d^{\prime}\sqrt{2\bar{m}V^{\prime}/\hbar^2}}.
\end{equation}
Instead of approximating the barrier shape by a rectangle we used
an inverted parabola. This gives the additional factor $\pi/4$ in
the exponent. The primes indicate that the values have been
changed by taking the harmonic ground level energies
$\frac{1}{2}\hbar (2w_2/m)^{0.5}$ of the relevant configurations,
into account, i.e. minima and saddle. This lowers the barrier height $V$ and decreases the
distance $d$. By using the values for $\Delta_0$, $E$ and
$P(\Delta_0, \Delta)$ and according to \eqnref{def_peff} with
$E/k_B$ limited to 1 K we obtain $P_{eff}=1.3\pm 0.5\cdot10^{46}
 \un{J}^{-1} \un{m}^{-3}$. This value is in good agreement
with previous results \cite{HeuerSilbey:1993a} and the
experimentally observed values for NiP \cite{Bellessa:1980} ($2\cdot10^{46}
J^{-1} m^{-3}$). This last value,
however, might be biased by the electron contribution in NiP and
is somewhat higher than commonly found values for other
glass-forming systems \cite{Berret:1988}, ranging from $0.5\cdot 10^{45}
\un{J}^{-1} \un{m}^{-3}$ to $3\cdot10^{45} \un{J}^{-1}
\un{m}^{-3}$. In disagreement with the \stm\
we observe an energy dependence of $\peff(\un{E})$ which can be
characterized as $\peff(E) \propto E^\delta_1$ with $\delta_1 =
0.14$, see Fig.4.

In the next step we compare the temperature dependence of the
thermal conductivity and the heat capacity calculated from
simulated \tls\ with experimentally found values; see Heuer and Silbey
\cite{HeuerSilbey:1993b} for the theoretical background. The
thermal conductivity is experimentally found to behave like
\begin{equation}
  \kappa(\un{T})\propto \frac{\un{T}^2}{\peff\gamma^2_{\sigma}}\propto \un{T}^{2-\beta}
\end{equation}
with $\beta$ small and positive. From this and
\begin{equation}
  \peff\propto \un{E}^{\delta_1}
\end{equation}
\begin{equation}
   \gamma_{\sigma}^2\propto d_{\un{eff}}^2(\un{E})\propto \un{E}^{\delta_2}
\end{equation}
follows $\beta=\delta_1+\delta_2$. $d_{\un{eff}}$ is defined as
\begin{equation}
  d_{eff}^2(E)=\frac{\sum_i^{\prime} d^2 \cdot \Delta_{i,0}^2/\un{E}_i^2}
                    {\sum_i^{\prime} \Delta_{i,0}^2/\un{E}_i^2},
\end{equation}
where the primes indicate that the sums include all \tls\ i with
an energy splitting smaller than E. The energy dependence of
$d_{eff}$ is shown in \figref{d2eff}, yielding
$\delta_2=-0.17$ and $\beta\approx 0.0$. For the calculation of
the heat capacity experimentally limited relaxation times have to
be taken into account. The heat capacity is experimentally found
to be
\begin{equation}
  \un{C(T)}\propto \un{T}^{1.1-1.3}.
\end{equation}
For a theoretical analysis one may use the expression
\begin{equation}
\label{eqc}
  \un{C(T,\tau_{exp})}\propto\int_0^\infty \frac{P_{\un{tot}}(E,\tau_{\un{exp}}) \un{E}^{2}}{4\un{k_B T}^2} sech\left[\frac{\un{E}}{2\un{K_B T}}\right]
     dE.
\end{equation}
Here $P_{\un{tot}}$ is the total number of \tls\ per energy and volume
with energy splitting below E and with relaxation times shorter
than $\tau_{\un{exp}}$. For a fixed relaxation time
$\tau_{\un{exp}}=5 s$ we obtain $P_{\un{tot}}(E,\tau_{\un{exp}})
\propto \un{E}^{1+\delta_3}$ with $\delta_3 = 0.21$; see Fig.4. The
value of $\delta_3$ is not very sensitive to the exact choice of
$\tau_{\un{exp}}$. Inserting this relation into Eq.\ref{eqc} one
obtains $\un{C(T,\tau_{exp})} \propto \un{T}^{1+\delta_3}$ in
agreement with the experiment.

\begin{figure}
  \includegraphics[width=\textwidth,clip]{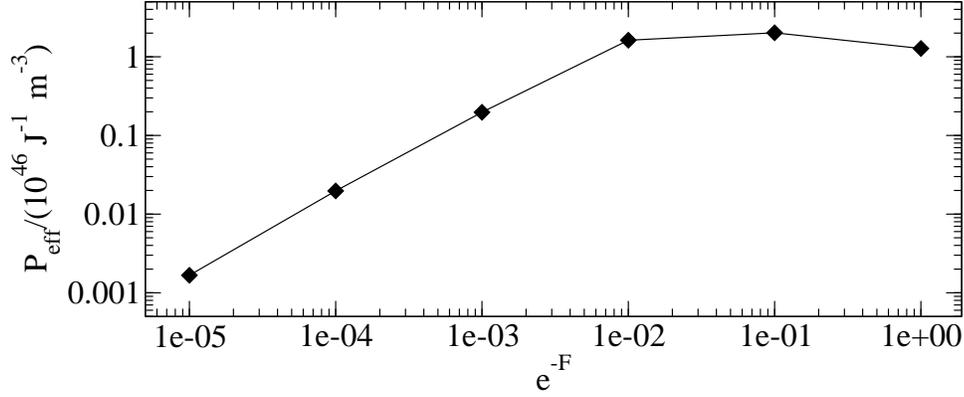}
  \caption{\label{peff_fc} Dependence of the \peff\ on the Franck-Condon factor.
            N=65, $E\approx$1 K}
\end{figure}
The comparison between experimentally determined results and
simulated results has to be done with care as the tunneling
matrix element is experimentally decreased by a mismatch of
the vibrational modes of the two states
giving rise to an additional factor \cite{Leggett:1987}.
\begin{equation}
  \Delta_{0, \un{exp}}=e^{-\un{F}}\cdot\Delta_0
\end{equation}
F is the Franck-Condon factor and does mainly depend on the
spectral density of eigenfrequencies and their coupling to the
system. The simulated number of \tls\ is therefore not directly
comparable with experimentally determined values. To check the
possible influence of $e^{-\un{F}}$ on our results some values, we have
calculated $\peff$ for different values of this factor; see
\figref{peff_fc}. Interestingly, \peff\ is slightly increased for
$0.01 < e^{-\un{F}} < 1$. The increase is caused by the fact that \tls\
with $E>1  K$ are lowered in their splitting energy such
that the  number of \tls\ below 1 K is increased. For
even smaller values of the Franck-Condon factor, however, the
$\Delta_0^2$-factor in Eq.\ref{def_peff} results in a significant
reduction of $\peff$.

\subsection{\label{r3}Microscopic Nature of the Two Level States}

\begin{figure}
  \includegraphics[width=\textwidth,clip]{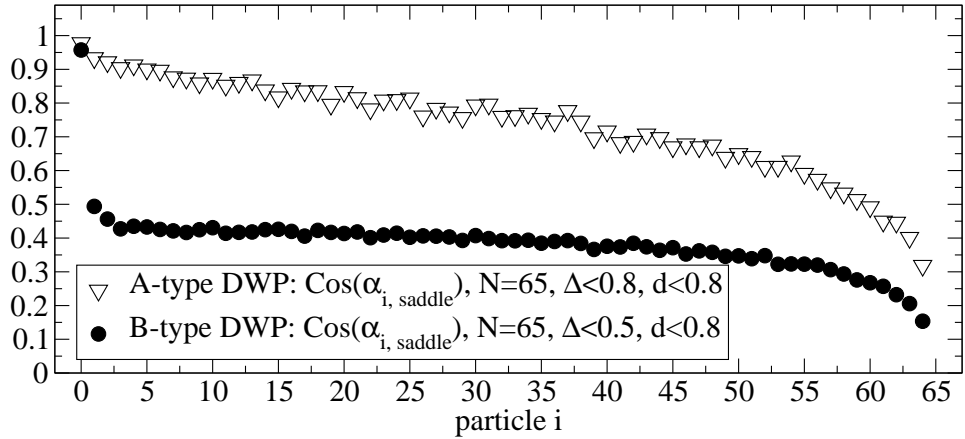}
  \caption{\label{drp_saddle_angle} Cosine of the angle between
           the transition vector  from the first minimum configuration to the saddle configuration and
       the vector from the saddle to the next minimum configuration. The particles
       are sorted according to their total motion (i=0: fastest particle). It is distinguished
        whether central particle is a A- (triangles) or a B-particle (spheres).}
\end{figure}

\begin{figure}
  \begin{center}
  \includegraphics[width=3.5in,clip]{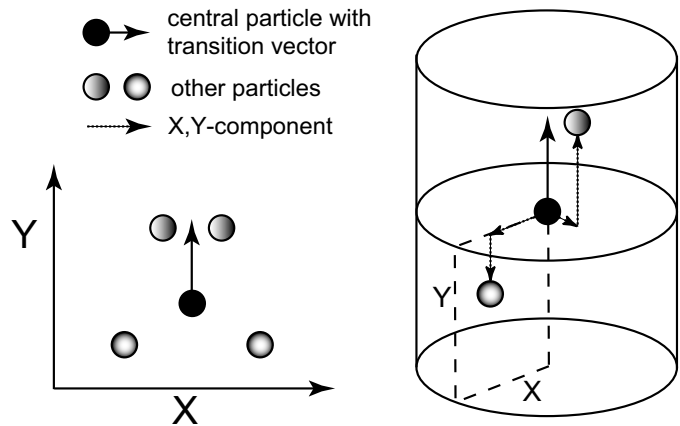}
  \caption{\label{d_definition}Sketch of the definition of the local
           environment of the central particle.}
  \end{center}
\end{figure}

\begin{figure}
  \includegraphics[width=\textwidth,clip]{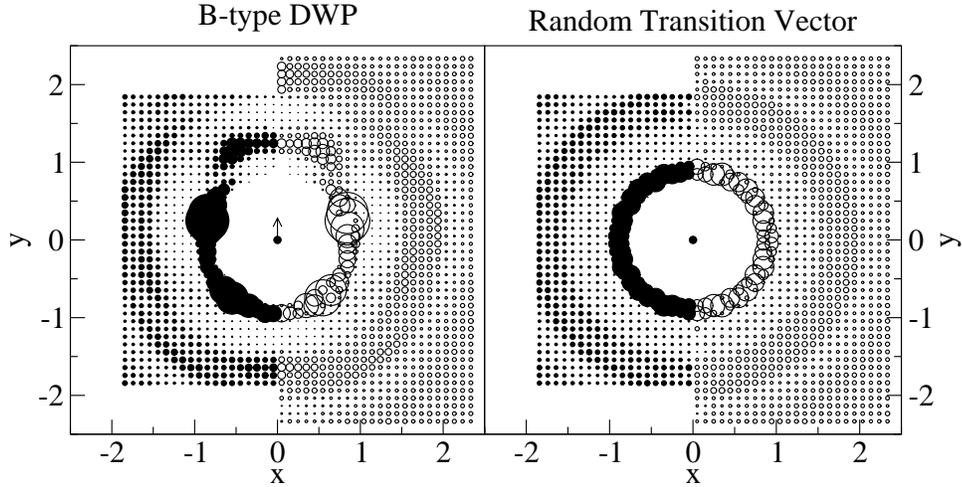}
  \caption{\label{density8_14_rand}Left: Averaged A-particle distribution
           around the central B particle. The filled circles are generated
       from a 65 particle system and the open circles are generated
       from a 130 particle system. There is no visible finite size effect.
           Right: Same as before but with a randomly chosen transition vector
       for the central particle. The density distribution corresponds
       to the radial distribution function.}
\end{figure}

\begin{figure}
  \includegraphics[width=\textwidth,clip]{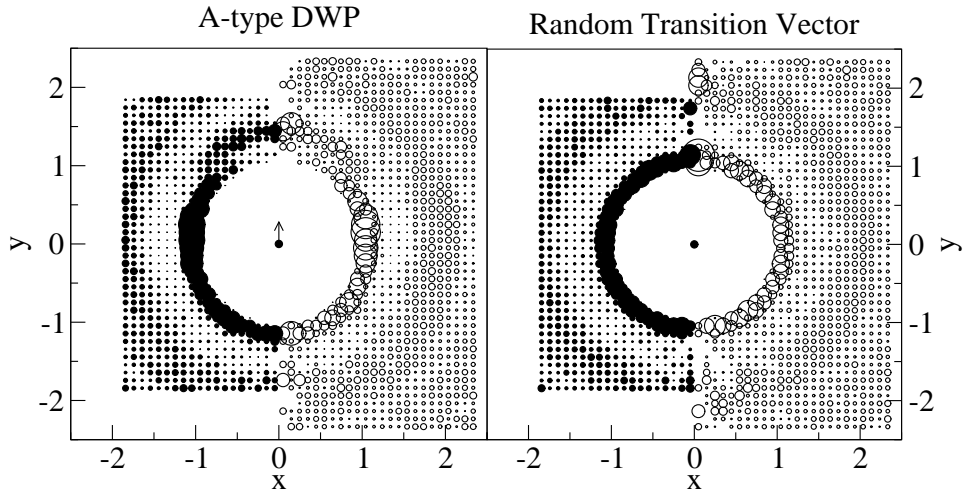}
  \caption{\label{density14_14_rand} Same as the previous figure except that central A-particles are considered.}
\end{figure}

  In our recent publication \cite{Reinisch_condmat:2004} we have analyzed
  the particle which shows
  the largest displacement during the transition between both minima. It is
  denoted {\it central} particle.
  \dwp\ with an A-particle as a central particle are denoted A-type
  \dwp. B-type \dwp\ are defined in analogy. Surprisingly, it turned
  out that 90\% of all \dwp\ are B-type \dwp\ although only 20\% of
  all particles are B-particle. Furthermore we see that for
  B-type \dwp\ the displacement of the central particles was significantly larger
  than that of all other particles. For A-type \dwp\ the distribution of displacements
  was more continuous. Stated differently, the transition in A-type \dwp\
  is more collective.

\begin{figure}
  \includegraphics[width=\textwidth,clip]{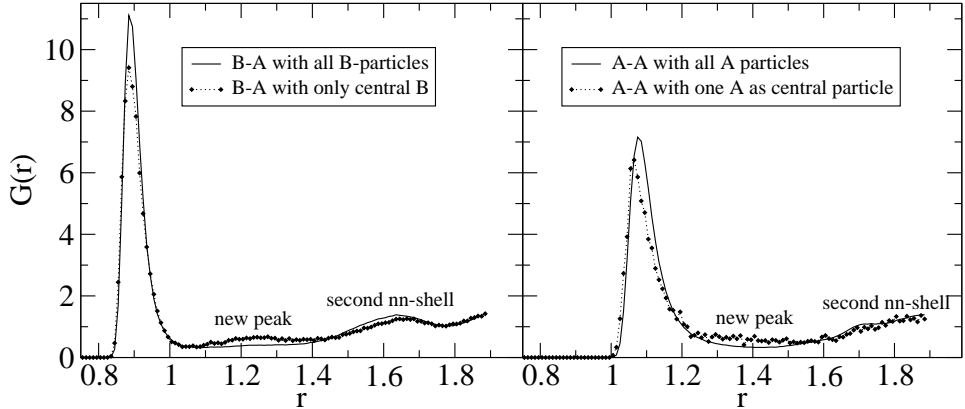}
  \caption{\label{gofr_p0} G(r) of the minimized structures
           in comparison to special G(r), where only the central particles in interaction
       with A-particles are considered.}
\end{figure}

\begin{figure}
  \includegraphics[width=\textwidth,clip]{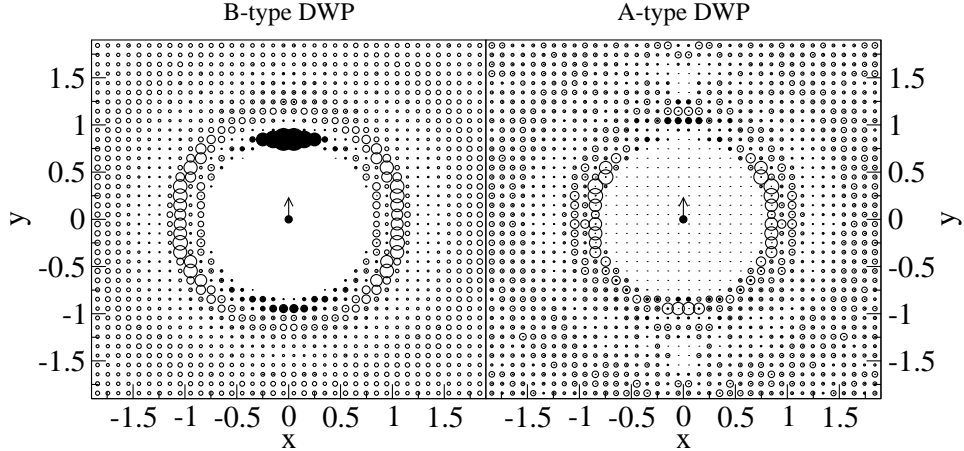}
  \caption{\label{density_second}Particle distribution around the particle
   with the second-largest displacement (N=65).
   In analogy to \figref{density8_14_rand} and \figref{density14_14_rand}
   the transition vector of this particle defines the y-axis.
   The filled circles mark the contribution from the central particle.}
\end{figure}
  In this subsection we elucidate the dynamics of \dwp\ much closer.
  First we analyze whether the displacement between both minima is along
  a straight line in configuration space or whether it is curved.
  More specifically
    we calculate the angle between the transition vector from the first minimum
    to the saddle and from the saddle to the second minimum; see \figref{drp_saddle_angle}.
    The results are sorted with the respect to the displacement of
    the different particles (i=0: particle with the largest
    displacement, i.e. central particle; i=64: particle with the
    smallest displacement).
  For B-type \dwp\ the central particle basically moves along a straight path
  whereas the other
  particles move along a strongly curved path. Thus the \dwp\ can be basically
  characterized as a one-particle motion of a B-particle. The other particles
  just support this transition by some
  complicated curved trajectory to optimize the energy.  The behavior is very different
  for A-type \dwp\ where most particles move along a relatively straight line.
  Since the central particle is only little different in terms of its
  displacement (see above) it is not surprising that
  the central particle behaves similarly as compared to the other particles.
  Obviously, the collective dynamics in A-type \dwp\ is realized by a rather
  straight translation of most particles.
  Note, however, that for both types of \dwp\ there is a tendency
  to move in a less curved way for particles with larger displacements, even if the
  reaction path approximation is considered.
  A priori this is not a necessity.

  In the next step we take a closer look into the local distribution of
  particles around the central particle. We therefore define the
  relative density in a cylinder around the central particle as shown
  in \figref{d_definition}. The transition vector of the central
  particle defines the y-axis. Furthermore the distance vectors between
  the central particles and the other particles in the
  starting minimum are represented in the xy-plane.
  By taking into account appropriate phase space factors the
  situation of an ideal gas would correspond to a homogeneous
  density distribution in the xy-plane. Finally, we average over
  all \dwp, distinguishing A-type and B-type \dwp. We only consider the
  density of A-particles around the central
  particles because the B-particles are of minor statistical relevance.
  \figref{density8_14_rand} shows the results for the B-type \dwp,
  which dominate the investigated BMLJ-system. The first striking
  observation is the distinct structure (left), which is
  different from the average pair correlation function (right).
  The main structural feature is the very small density
  for $(\un{x}\approx 0, \un{y} \approx   \sigma_{\un{AB}} = 0.88)$.
  This directly shows that the central B-particles moves in a direction
  where free space
  is available.
  An increased density in the nearest neighbor shell (nn-shell) is found
  orthogonal to the direction of the translation vector.
  Thus the central particle jumps
  through this structure to get to the new minimum. Note that
  results for N=65 and N=130 are identical within numerical
  uncertainties. For the larger system one can see that also the
  second nearest neighbor shell reflects the properties of nn-shell.
  On a qualitative basis the same properties are observed for A-type \dwp;
  see \figref{density14_14_rand}.
  It is, however, much less pronounced. This is a consequence of the
  fact that A-type \dwp\ are more collective.

  A priori it is not clear whether this observation implies an
  anomalous structure around the central particle. This is checked by analyzing
  the radial distribution function G(r) for the average minimized structures and
  G(r) for the central particles of the minimized structures, see \figref{gofr_p0}.
  We can clearly see that for both types of \dwp\ density is taken from the first
  nn-shell and transferred to a new peak between the first and the
  second nearest neighbor shell. Thus the observed structure in the local environment
  does not arise from the fact that the central transition vector chooses
  a special direction in an average environment, but that the environment
  is indeed altered.

  To illustrate the above observations we show a selected \dwp\ which explicitly displays
  the  structural patterns; see \figref{dwp_illustration}.
  Please note the particles in the intermediate shell, as well as the particles orthogonal
  to the transition vector.

\begin{figure}
  \includegraphics[width=\textwidth,clip]{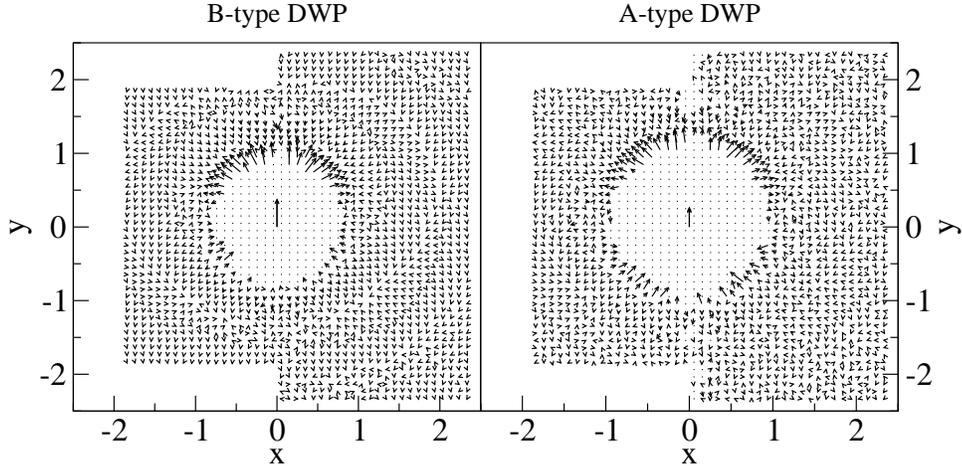}
  \caption{\label{vector} Averaged projected displacement of all A-particles
           relative to the transition of the central particle.
       Results for N=65 and N=130 are shown.}
\end{figure}
\begin{figure}
  \includegraphics[width=\textwidth,clip]{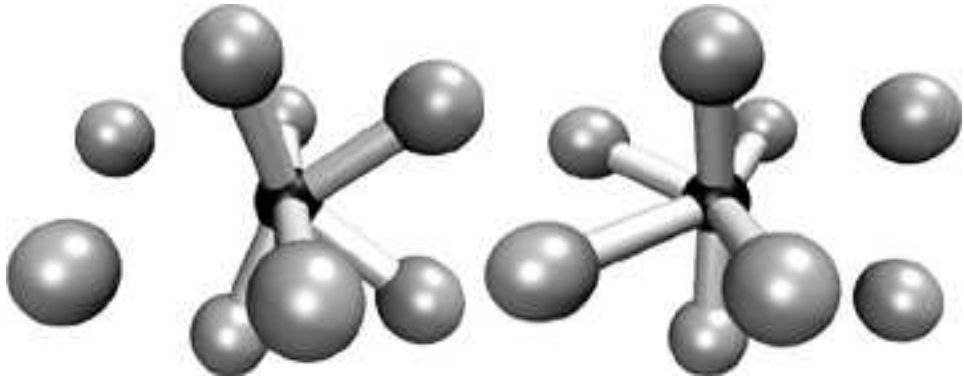}
  \caption{\label{dwp_illustration} Type B-\dwp\ as observed from the N=65 simulation.
           The small black central particle is a B-particle. The shown bonds have
       a length around 0.9 the non bonded atoms have a distance to the B-particle
       between 1.11 and 1.21. The two configurations correspond to the
       minima of a \dwp.}
\end{figure}
  To go further into structural details we repeat the analysis, performed so far for
  the central particle, also for the particle with the second-largest displacement.
  The data are presented in \figref{density_second} for both types of \dwp.
  The structural changes for the B-type \dwp\ are dramatic if
  compared to \figref{density8_14_rand}. For the B-type \dwp\ the first shell
  is almost back to normal except for an increased probability to find particles
  just in front of the second particle. This density is
  mainly caused by the central particle. This means that the particle with
  the second-largest displacement mainly follows the central B-particle.
  The A-type \dwp\ still shows a similar picture as \figref{density14_14_rand}(left)
  but much less pronounced, the central particle shows only a small tendency
  to occupy positions in front of the second particle (tiny filled circles
  ahead of the second particle).
  This again is a direct consequence of the fact that for A-type
  \dwp\ the central particle is not behaving very different
  to the other participating particles.

So far we have analyzed the structure around the central particle.
In the final step we also  elucidate the displacement of the
particles  around the central particle. The representation in
\figref{vector} is analog
  to those in \figref{d_definition}, the x-y-plane is still the same, but
  now we do not analyze where particles are in the plane but how they move
  relative to the central transition vector. The displacement is projected into the
  x-y-plane. Both graphs in \figref{vector} look similar, the differences between
  A- and B-type \dwp\ are on a quantitative level.
  The general explanation for the observed
  displacements are simple. The fastest particle makes a jump from
  one minimum to the other and all other particles follow as if
 they were connected to the central particle by a spring, reflecting the attractive
 part of the Lennard-Jones potential.

\section{SUMMARY}
  We were able to present a huge set of \dwp, which were found on a systematic basis.
The number of \tls\ is consistent with experimental observations
and the deviations from the predictions of the \stm\ can be
explained in terms of the energy dependence of \peff\ and \ptot.
These results were already part of older work
\cite{HeuerSilbey:1993b} but are now reproduced on a much better
statistical basis.

  With the analysis of the saddle angles during the transition and the structure
  of the second particle we underline the single particle nature of the transition
  in  B-type \dwp, as already proposed in \cite{Reinisch_condmat:2004}.
  As the B-type \dwp\ are predominant in the investigated BMLJ-system
  it follows that single particle-type \dwp\ determine the low temperature
  properties of BMLJ.
  The more collective A-type transitions are not as important for our system
  but can be used as a model for
  other substances, where all molecules have approximately the same size
  and mobility.

  We were also able to present a detailed analysis of the structure
  and the displacements of the \dwp. It is shown that the presence of holes
  in the structure is a prerequisite for the formation of \dwp. For B-type \dwp\
  one basically has a one-particle displacement. The other particles mainly
  follow the central particle. In contrast, for A-type \dwp\ the
  dynamics are much more collective.

In a next step we aim to compare the properties with that of
network glasses like silicate in order to see how the structural
differences are reflected in the low-temperature properties of
glasses.

\section*{ACKNOWLEDGMENTS}
We like to thank B. Doliwa, H. R. Schober and G. Viliani for fruitful discussions and
the International Graduate School of Chemistry for funding.
\bibliographystyle{unsrt}
\bibliography{ljtt2_v4}

\end{document}